# THE BASICS OF LINE MOIRÉ PATTERNS AND OPTICAL SPEEDUP

*Emin Gabrielyan*
Switzernet Sàrl, Scientific Park of Swiss Federal
Institute of Technology, Lausanne (EPFL)
emin.gabrielyan@switzernet.com

## Abstract

*We are addressing the optical speedup of movements of layers in moiré patterns. We introduce a set of equations for computing curved patterns, where the formulas of optical speedup and moiré periods are kept in their simplest form. We consider linear movements and rotations. In the presented notation, all periods are relative to the axis of movements of layers and moiré bands.*

## 1. Introduction

Moiré patterns appear when superposing two transparent layers containing correlated opaque patterns. The case when layer patterns comprise straight or curved lines is called line moiré.

This document presents the basics of line moiré patterns. We present numerous examples and we focus also on the optical speedup of moiré shapes when moving layer patterns. Numerous examples are present. Dynamic examples demonstrating the movements of layers are presented by GIF files (hyperlinks are provided in square brackets).

We develop here the most important formulas for computing the periods of superposition patterns, the inclination angles and the velocities of optical shapes when moving one of the layers.

In section 2, we demonstrate the phenomenon on the examples with horizontal parallel lines, which are further extended to cases with inclined and curved lines. In section 3 we present circular examples with straight radial lines, which are analogously extended.

## 2. Simple moiré patterns

### 2.1. Superposition of layers with periodically repeating parallel lines

Simple moiré patterns can be observed when superposing two transparent layers comprising periodically repeating opaque parallel lines as shown in Figure 1. The lines of one layer are parallel to the lines of the second layer.

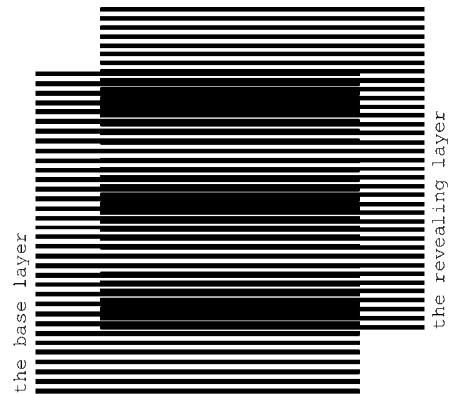

**Figure 1. Superposition of two layers consisting of parallel lines, where the lines of the revealing layer are parallel to the lines of the base layer** [eps], [tif], [png]

The superposition image does not change if transparent layers with their opaque patterns are inverted. We denote one of the layers as the *base layer* and the other one as the *revealing layer.* When considering printed samples, we assume that the revealing layer is printed on a transparency and is superposed on top of the base layer, which can be printed either on a transparency or on an opaque paper. The periods of the two layer patterns, i.e. the space between the axes of parallel lines, are close. We denote the period of the base layer as $p_b$ and the period of the revealing layer as $p_r$. In Figure 1, the period of lines of the base layer is equal to 6 units, and the period of lines of the revealing layer is equal to 5.5 units.

The superposition image of Figure 1 outlines periodically repeating dark parallel bands, called moiré lines. Spacing between the moiré lines is much larger than the periodicity of lines in the layers.

Light areas of the superposition image correspond to the zones where the lines of both layers overlap. The dark areas of the superposition image forming the moiré lines correspond to the zones where the lines of the two layers interleave, hiding the white background. The labels of Figure 2 show the passages from light zones with overlapping layer lines to dark zones with interleaving layer lines. The light and dark zones are periodically interchanging.

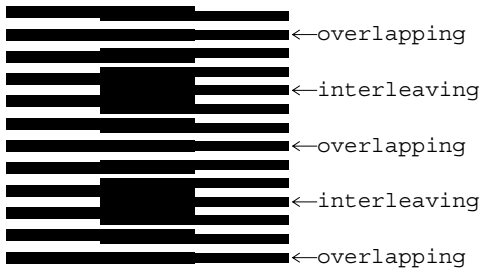

**Figure 2. Superposition of two layers consisting of horizontal parallel lines [eps], [tif], [png]**

Figure 3 shows a detailed diagram of the superposition image between two light zones, where the lines of the revealing and base layers overlap [Sciammarella62a, p. 584].

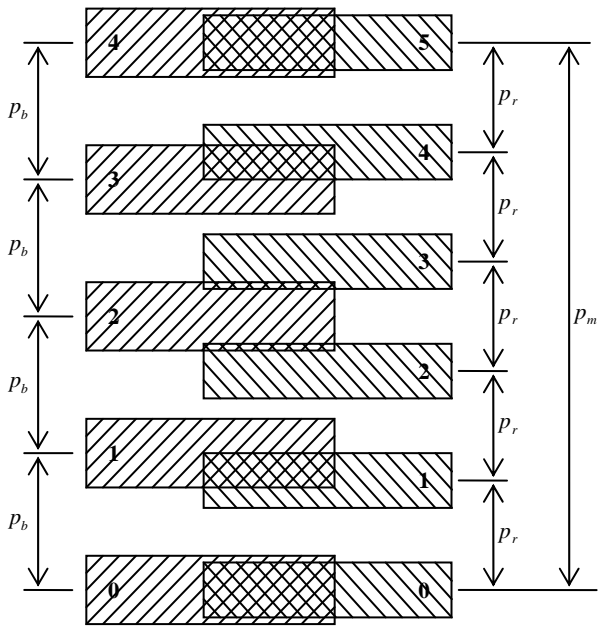

**Figure 3. Computing the period of moiré lines in a superposition image as a function of the periods of lines of the revealing and base layers**

The period $p_m$ of moiré lines is the distance from one point where the lines of both layers overlap (at the bottom of the figure) to the next such point (at the top). Let us count the layer lines, starting from the point where they overlap. Since in our case $p_r < p_b$, for the same number of counted lines, the base layer lines with a long period advance faster than the revealing layer lines with a short period. At the halfway of the distance $p_m$, the base layer lines are ahead the revealing layer lines by a half a period ($p_r/2$) of the revealing layer lines, due to which the lines are interleaving, forming a dark zone. At the full distance $p_m$, the base layer lines are ahead of the revealing layer lines by a full period $p_r$, so the lines of the layers again overlap. The base layer lines gain the distance $p_m$ with as many lines ($p_m/p_b$) as the number of the revealing layer lines ($p_m/p_r$) for the same distance minus one:

$$\frac{p_m}{p_r} = \frac{p_m}{p_b} + 1 \qquad (2.1)$$

From equation (2.1) we obtain the well known formula for the period $p_m$ of the superposition image [Amidror00a, p.20]:

$$p_m = \frac{p_b \cdot p_r}{p_b - p_r} \qquad (2.2)$$

The superposition of two layers comprising parallel lines forms an optical image comprising parallel moiré lines with a magnified period. According to equation (2.2), the closer the periods of the two layers, the stronger the magnification factor is.

If the numbers $p_m/p_b$ and $p_m/p_r$ are integers, then if at some moiré light zone the lines of both layers perfectly overlap, as shown in Figure 3, the layer lines will also perfectly overlap also at the centers of each other light zone. If $p_m/p_b$ and $p_m/p_r$ are not integers, then the centers of white moiré zones do not necessarily match with the centers of layer lines. In any case, equation (2.2) remains valid.

For the case when the revealing layer period is longer than the base layer period, the space between moiré lines of the superposition pattern is the absolute value of formula of (2.2).

The thicknesses of layer lines affect the overall darkness of the superposition image and the thickness of the moiré lines, but the period $p_m$ does not depend on the layer lines' thickness. In our examples the base layer lines' thickness is equal to $p_b/2$, and the revealing layer lines' thickness is equal to $p_r/2$.

### 2.2. Speedup of movements with moiré

The moiré bands of Figure 1 will move if we displace the revealing layer. When the revealing layer moves perpendicularly to layer lines, the moiré bands move along the same axis, but several times faster than the movement of the revealing layer.

The three images of Figure 4 show the superposition image for different positions of the revealing layer. In the second image (b) of Figure 4, compared to the first image (a), the revealing layer is shifted up by one third of the revealing layer period ($p_r/3$). In the third image (c), compared to the first image (a), the revealing layer is shifted up by two third of the revealing layer period ($p_r \cdot 2/3$). The images show that the moiré lines of the superposition image move up at a speed, much faster than the speed of movement of the revealing layer.

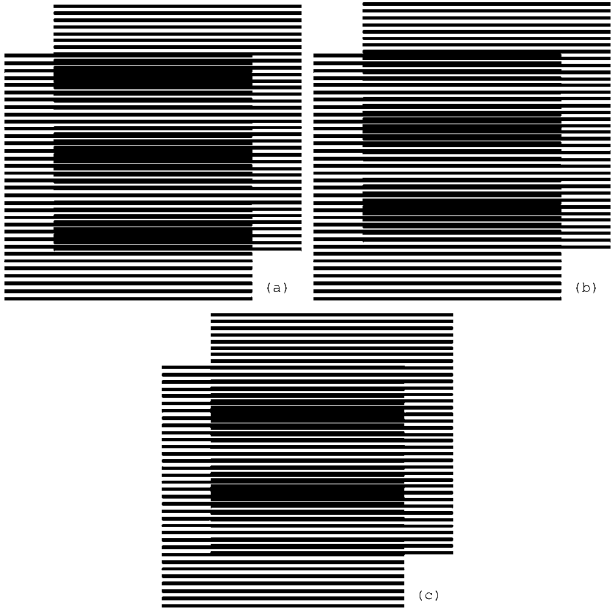

**Figure 4. Superposition of two layers with parallel horizontal lines, where the revealing layer moves vertically at a slow speed [eps (a)], [png (a)], [eps (b) ], [png (b)], [eps (c)], [png (c)]**

When the revealing layer is shifted up perpendicularly to the layer lines by one full period ($p_r$) of its pattern, the superposition optical image must be the same as the initial one. It means that the moiré lines traverse a distance equal to the period of the superposition image $p_m$ while the revealing layer traverses the distance equal to its period $p_r$. Assuming that the base layer is immobile ($v_b = 0$), the following equation holds for the ratio of the optical image's speed to the revealing layer's speed:

$$\frac{v_m}{v_r} = \frac{p_m}{p_r} \qquad (2.3)$$

According to equation (2.2) we have:

$$\frac{v_m}{v_r} = \frac{p_b}{p_b - p_r} \qquad (2.4)$$

In case the period of the revealing layer is longer than the period of the base layer, the optical image moves in the opposite direction. The negative value of the ratio computed according to equation (2.4) signifies the movement in the reverse direction.

The GIF animation of the superposition image corresponding to a slow movement of the revealing layer is provided [ps], [gif], [tif]. The GIF file repeatedly animates a perpendicular movement of the revealing layer across a distance equal to $p_r$.

## 2.3. Superposition of layers with inclined lines

In this section we develop equations for patterns with inclined lines. Since most of all we are interested in optical speedup, instead of using the well known equations, we represent the case of inclined patterns such that the equations (2.2), (2.3), and (2.4) remain valid in their current simple form. The values of periods $p_r$, $p_b$, and $p_m$ for the examples of Figure 4 correspond to the distances between the lines along the vertical axis corresponding to the axis of movements. When the layer lines are horizontal (or perpendicular to the movement axis) the periods (*p*) are equal to the distances (denoted as *T*) between the lines (as in Figure 1, Figure 3, and Figure 4). If the lines are inclined the periods (*p*) along the vertical axis does not correspond anymore to the distances (*T*) between the lines. According to our notations, the periods *p* do not represent the distances *T* between the lines, but the distances between the lines along the axis of movements. By adopting the new notation, equations (2.2), (2.3), and (2.4) are valid all the time. Equations for inclination angles for such notation of periods (*p*) are presented in this section. For rotational movements *p* values represent the periods along circumference, i.e. the angular periods.

### 2.3.1. Computing moiré lines' inclination as function of the inclination of layers' lines

The superposition of two layers with identically inclined lines forms moiré lines inclined at the same angle. Figure 5 is obtained from Figure 1 with a vertical shearing. In Figure 5 the layer lines and the moiré lines are inclined by 10 degrees. Inclination is not a rotation. During the inclination the distance between the layer lines along the vertical axis ($p$) is conserved, but the true distance *T* between the lines (along an axis perpendicular to these lines) changes. The diagram of Figure 8 shows the difference between the vertical periods $p_b$ and $p_r$, and the distances $T_b$ and $T_r$.

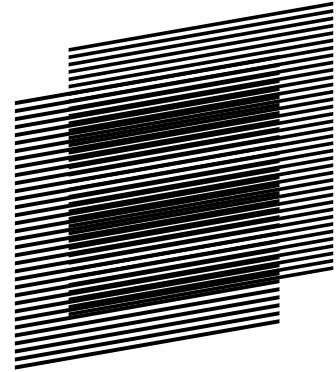

**Figure 5. Superposition of layers consisting of inclined parallel lines where the lines of the base and revealing layers are inclined at the same angle [eps], [png]**

The inclination degree of layer lines may change along the horizontal axis forming curves. The superposition of two layers with identical inclination pattern forms moiré curves with the same inclination pattern. In Figure 6 the inclination degree of layer lines gradually changes according the following sequence of degrees (+30, –30, +30, –30, +30), meaning that the curve is divided along the horizontal axis into four equal intervals and in each such interval the curve's inclination degree linearly changes from one degree to the next according to the sequence of five degrees. Layer periods $p_b$ and $p_r$ represent the distances between the curves along the vertical axis. In Figure 5 and Figure 6, $p_b$ is equal to 6 units and $p_r$ is 5.5. units. Figure 6 can be obtained from Figure 1 by interpolating the image along the horizontal axis into vertical bands and by applying a corresponding vertical shearing and shifting to each of these bands. Equation (2.2) is valid for computing the spacing $p_m$ between the moiré curves along the vertical axis and equation (2.4) for computing the optical speedup ratio when the revealing layer moves along the vertical axis.

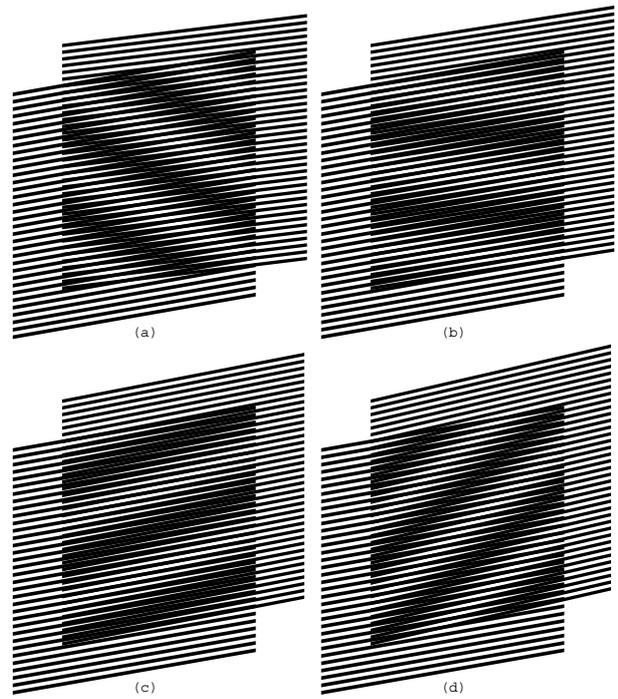

Figure 7. Superposition of layers consisting of inclined parallel lines, where the base layer lines' inclination is 10 degrees and the revealing layer lines' inclination is 7, 9, 11, and 13 degrees [eps (a)], [png (a)], [eps (b)], [png (b)], [eps (c)], [png (c)], [eps (d)], [png (d)]

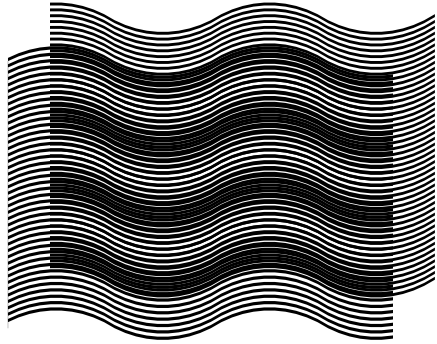

Figure 6. Two layers consisting of curves with identical inclination patterns, and the superposition image of these layers [eps], [png]

More interesting is the case when the inclination degrees of layer lines are not the same for the base and revealing layers. Figure 7 shows four superposition images where the inclination degree of base layer lines is the same for all images (10 degrees), but the inclination of the revealing layer lines is different for images (a), (b), (c), and (d) and is equal to 7, 9, 11, and 13 degrees correspondingly. The periods of layers along the vertical axis $p_b$ and $p_r$ (6 and 5.5 units correspondingly) are the same for all images. Correspondingly, the period $p_m$ computed with equation (2.2) is also the same for all images.

We provide a GIF animation of the superposition image when the revealing layer's inclination oscillates between 5 and 15 degrees [ps], [gif], [tif].

Figure 8 helps to compute the inclination degree of moiré optical lines as a function of the inclination of the revealing and the base layer lines. We draw the layer lines schematically without showing their true thicknesses. The bold lines of the diagram inclined by $\alpha_b$ degrees are the base layer lines. The bold lines inclined by $\alpha_r$ degrees are the revealing layer lines. The base layer lines are vertically spaced by a distance equal to $p_b$, and the revealing layer lines are vertically spaced by a distance equal to $p_r$. The distances $T_b$ between the base layer lines and the distance $T_r$ between the revealing layer lines are not used for the development of the next equations. The intersections of the lines of the base and the revealing layers (marked in the figure by two arrows) lie on a central axis of a light moiré band between dark moiré lines. The dashed line of Figure 8 corresponds to the axis of the light moiré band between two moiré lines. The inclination degree of moiré lines is therefore the inclination $\alpha_m$ of the dashed line.

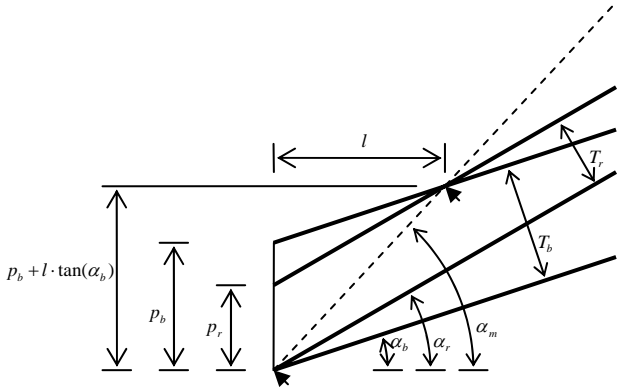

**Figure 8.** Computing the inclination angle of moiré lines as a function of inclination angles of the base layer and revealing layer lines

From Figure 8 we deduce the following two equations:

$$\begin{cases} \tan \alpha_m = \dfrac{p_b + l \cdot \tan \alpha_b}{l} \\ \tan \alpha_r = \dfrac{p_b - p_r + l \cdot \tan \alpha_b}{l} \end{cases} \quad (2.5)$$

From these equations we deduce the equation for computing the inclination of moiré lines as a function of the inclinations of the base layer and the revealing layer lines:

$$\tan \alpha_m = \frac{p_b \cdot \tan \alpha_r - p_r \cdot \tan \alpha_b}{p_b - p_r} \quad (2.6)$$

For the base layer inclination fixed to 30 degree, with a base layer period equal to 6 units, and with a revealing layer period equal to 5.5 units, the bold curve of Figure 9 represents the moiré line inclination degree as a function of the revealing layer line inclination. The two other curves correspond to cases, when the base layer inclination is equal to 20 and 40 degrees correspondingly. The circle marks correspond to the points where both layers' lines inclinations are equal, and the moiré lines inclination also become the same.

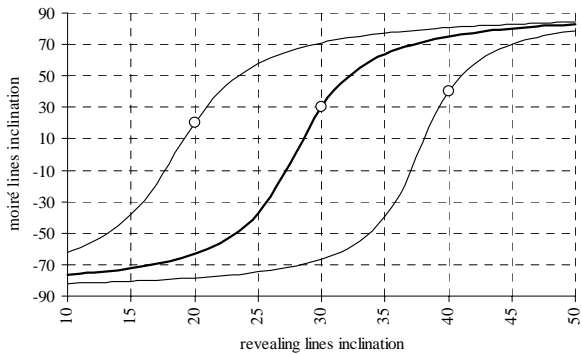

**Figure 9.** Moiré lines inclination as a function of the revealing layer lines inclination for the base layer lines inclination equal to 30 degrees [xls]

### 2.3.2. Deducing the known formulas from our equations

The periods $T_b$, $T_r$, and $T_m$ used in the literature are computed as follows (see Figure 8):

$$T_b = p_b \cdot \cos \alpha_b, \; T_r = p_r \cdot \cos \alpha_r, \\ T_m = p_m \cdot \cos \alpha_m \quad (2.7)$$

From here, using equation (2.6) we deduce the well known formula for the angle of moiré lines [Amidror00a]:

$$\alpha_m = \arctan\left(\frac{T_b \cdot \sin \alpha_r - T_r \cdot \sin \alpha_b}{T_b \cdot \cos \alpha_r - T_r \cdot \cos \alpha_b}\right) \quad (2.8)$$

Recall from trigonometry the following simple formulas:

$$\cos \alpha = \frac{1}{\sqrt{1 + \tan^2 \alpha}} \quad (2.9)$$

$$\cos(\alpha_1 - \alpha_2) = \cos \alpha_1 \cdot \cos \alpha_2 + \sin \alpha_1 \cdot \sin \alpha_2$$

From equations (2.8) and (2.9) we have:

$$\cos \alpha_m = \frac{T_b \cdot \cos \alpha_r - T_r \cdot \cos \alpha_b}{\sqrt{T_b^2 + T_r^2 - 2 \cdot T_b \cdot T_r \cdot \cos(\alpha_r - \alpha_b)}} \quad (2.10)$$

From equations (2.2) and (2.7) we have:

$$T_m = \frac{T_b \cdot T_r}{T_b \cdot \cos \alpha_r - T_r \cdot \cos \alpha_b} \cdot \cos \alpha_m \quad (2.11)$$

From equations (2.10) and (2.11) we deduce the second well known formula for the period $T_m$ of moiré lines:

$$T_m = \frac{T_b \cdot T_r}{\sqrt{T_b^2 + T_r^2 - 2 \cdot T_b \cdot T_r \cdot \cos(\alpha_r - \alpha_b)}} \quad (2.12)$$

Recall from trigonometry that:

$$\sin \frac{\alpha}{2} = \sqrt{\frac{1 - \cos \alpha}{2}} \quad (2.13)$$

In the particular case when $T_b = T_r$, taking in account equation (2.13), equation (2.12) is further reduced into well known formula:

$$T_m = \frac{T}{2 \cdot \sin\left(\dfrac{\alpha_r - \alpha_b}{2}\right)} \quad (2.14)$$

Still for the case when $T_b = T_r$, we can temporarily assume that all angles are relative to the base layer lines and rewrite equation (2.8) as follows:

$$\alpha'_m = \arctan\left(\frac{\sin \alpha'_r}{\cos \alpha'_r - 1}\right) \quad (2.15)$$

Recall from trigonometry that:

$$\tan \frac{\alpha}{2} = \frac{1 - \cos \alpha}{\sin \alpha} \\ \tan(90° + \alpha) = -\frac{1}{\tan \alpha} \quad (2.16)$$

Therefore from equations (2.15) and (2.16):

$$\alpha'_m = 90° + \frac{\alpha'_r}{2} \quad (2.17)$$

Now for the case when the revealing layer lines do not represent the angle zero:

$$\alpha_m = \alpha_b + 90° + \frac{\alpha_r - \alpha_b}{2} \quad (2.18)$$

We obtain the well known formula [Amidror00a]:

$$\alpha_m = 90° + \frac{\alpha_r + \alpha_b}{2} \quad (2.19)$$

Equations (2.8) and (2.12) are the general case formulas known in the literature, and equations (2.14) and (2.19) are the formulas for rotation of identical patterns with parallel lines (i.e. the case when $T_b = T_r$) [Amidror00a], [Nishijima64a], [Oster63a], [Morse61a].

Assuming in the well known equation (2.8) that $\alpha_b = 0$, Figure 10 shows the charts of the moiré lines' degree as a function of the revealing layer's rotation degree for different values of $T_r / T_b$.

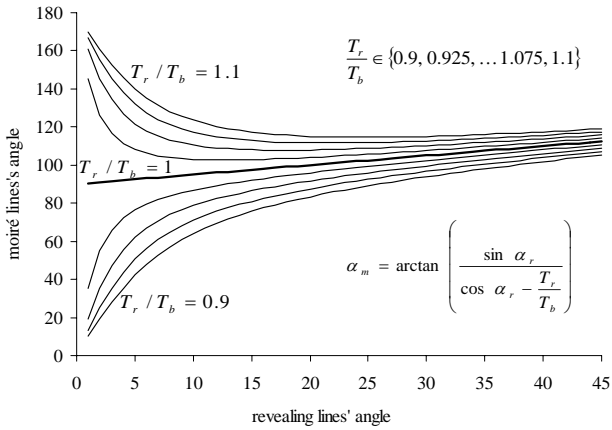

**Figure 10. Moiré lines inclination as a function of the rotation degree of the revealing layer [xls]**

Only for the case when $T_b = T_r$ (the bold curve) the rotation of moiré lines is linear with respect to the rotation of the revealing layer. Comparisons of Figure 10 and Figure 9 show the significant difference between shearing (i.e. inclination of lines) and rotation of the revealing layer pattern.

### 2.3.3. The revealing lines inclination as a function of the superposition image's lines inclination

From equation (2.6) we can deduce the equation for computing the revealing layer line inclination $\alpha_r$ for a given base layer line inclination $\alpha_b$, and a desired moiré line inclination $\alpha_m$:

$$\tan \alpha_r = \frac{p_r}{p_b} \cdot \tan \alpha_b + \left(1 - \frac{p_r}{p_b}\right) \cdot \tan \alpha_m \quad (2.20)$$

The increment of the tangent of revealing lines' angle ($\tan(\alpha_r) - \tan(\alpha_b)$) relatively to the tangent of the base layer lines' angle can be expressed, as follows:

$$\tan \alpha_r - \tan \alpha_b = \left(1 - \frac{p_r}{p_b}\right) \cdot (\tan \alpha_m - \tan \alpha_b) \quad (2.21)$$

According to equation (2.4), $1 - p_r / p_b$ is the inverse of the optical acceleration factor, and therefore equation (2.21) can be rewritten as follows:

$$\frac{\tan \alpha_r - \tan \alpha_b}{\tan \alpha_m - \tan \alpha_b} = \frac{v_r}{v_m} \quad (2.22)$$

Equation (2.22) shows that relative to the tangent of the base layer lines' angle, the increment of the tangent of the revealing layer lines' angle needs to be smaller than the increment of the tangent of the moiré lines' angle, by the same factor as the optical speedup.

For any given base layer line inclination, equation (2.20) permits us to obtain a desired moiré line inclination by properly choosing the revealing layer inclination. In Figure 6 we showed an example where the curves of layers follow an identical inclination pattern forming a superposition image with the same inclination pattern. The inclination degrees of the layers' and moiré lines change along the horizontal axis according the following sequence of alternating degree values (+30, –30, +30, –30, +30). In Figure 11 we obtained the same superposition pattern as in Figure 6, but the base layer consists of straight lines inclined by –10 degrees. The revealing pattern of Figure 11 is computed by interpolating the curves into connected straight lines, where for each position along the horizontal axis, the revealing line's inclination angle is computed as a function of $\alpha_b$ and $\alpha_m$, according to equation (2.20). Figure 11 demonstrates what is already expressed by equation (2.22): the difference between the inclination patterns of the revealing layer and the base layer are several times smaller than the difference between the inclination patterns of moiré lines and the base layer lines.

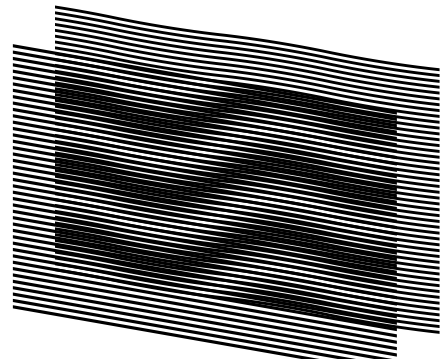

**Figure 11. The base layer with inclined straight lines, the revealing layer computed so as to form the desired superposition image [eps], [png]**

Another example forming the same superposition patterns as in Figure 6 and Figure 11 is shown in Figure 12. Note that in Figure 12 the desired inclination pattern (+30, –30, +30, –30, +30) is obtained using a base layer with an inverted inclination pattern (–30, +30, –30, +30, –30).

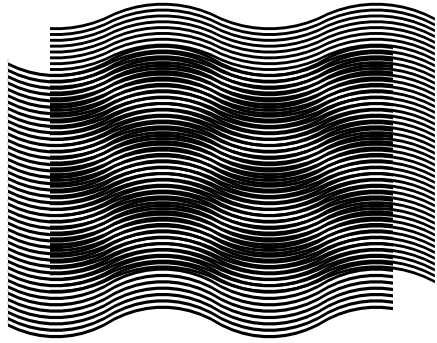

**Figure 12. A superposition image, where the base layer and moiré curves are mirrored relatively to the horizontal axis [eps], [png]**

We provide a GIF animation where we show a superposition image with a constant inclination pattern of moiré lines (+30, –30, +30, –30, +30) for modifying pairs of base and revealing layers [ps], [tif], [gif]. The base layer inclination pattern gradually changes and the revealing layer inclination pattern correspondingly adapts such that the superposition image's inclination pattern remains the same.

# 3. Superposition of periodic circular patterns

## 3.1. Superposition of circular periodic patterns with radial lines

Similarly to layer and moiré patterns comprising parallel lines (see Figure 1, Figure 2, and Figure 3), concentric superposition of dense periodic layer patterns comprising radial lines forms magnified periodic moiré patterns also with radial lines.

Figure 13 is the counterpart of Figure 1, where the horizontal axis is replaced by the radius and the vertical axis by the angle. Full circumferences of layer patterns are equally divided by integer numbers of radial lines. The number of radial lines of the base layer is denoted as $n_b$ and the number of radial lines of the revealing layer is denoted as $n_r$.

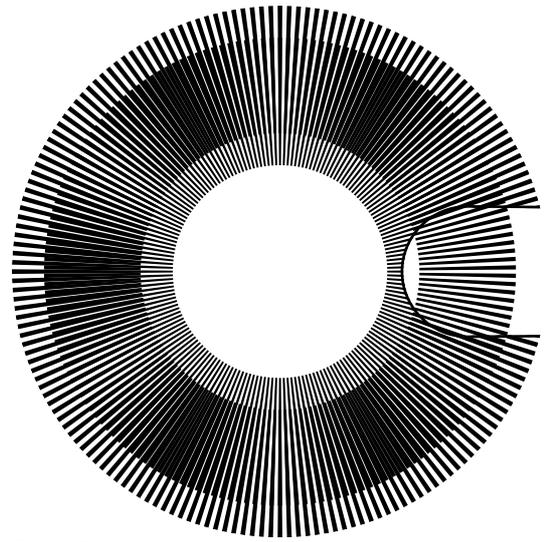

**Figure 13. Superposition of two layers with regularly spaced radial segments (a portion of the revealing layer is cut out to show a part of the base layer in the background) [eps], [png], [ps], [tif], [gif], small version [eps], [png]**

The periods $p_b$ and $p_r$ denote the angles between the central radial axes of adjacent lines. Therefore:

$$p_b = \frac{360°}{n_b}, \quad p_r = \frac{360°}{n_r} \qquad (3.1)$$

According to equations (3.1), equation (2.2) can be rewritten as follows:

$$p_m = \frac{360°}{n_r - n_b} \qquad (3.2)$$

Therefore the number of moiré radial lines $n_m$ corresponds to the difference between the numbers of layer lines:

$$n_m = |n_r - n_b| \qquad (3.3)$$

If in the layer patterns, the full circumferences are divided by integer numbers of layer lines, the circumference of the superposition image is also divided by an integer number of more lines.

Radial lines in Figure 13 have constant angular thickness, giving them the forms of segments, thick at their outer ends and thin at their inner ends. The thickness of radial lines affects the overall darkness of the superposition image and the width of moiré bands, but there is no impact on other factors, such as period of superposition pattern (i.e. values of $p_m$ and $n_m$). In our examples the angular thicknesses of layer lines are equal to the layer's half-period, i.e. the thickness of the base layer lines is equal to $p_b/2$ and the thickness of the revealing layer lines is $p_r/2$.

The optical speedup factor of equation (2.4) can be rewritten by replacing the periods $p_r$ and $p_b$ by their expressions from equations (3.1):

$$\frac{v_m}{v_r} = \frac{n_r}{n_r - n_b} \qquad (3.4)$$

The values $v_r$ and $v_m$ represent the angular speeds. The negative speedup signifies a rotation of the superposition image in a direction inverse to the rotation of the revealing layer. Considering (3.3), the absolute value of the optical speedup factor is:

$$\left|\frac{v_m}{v_r}\right| = \frac{n_r}{n_m} \qquad (3.5)$$

In Figure 13, the number of radial lines of the revealing layer is equal to 180, and the number of radial lines of the base layer is 174. Therefore, according to equations (3.4) and (3.3), the optical speedup is equal to 30, confirmed by the two images (a) and (b) of Figure 14, and the number of moiré lines is equal to 6, confirmed by the image of Figure 13.

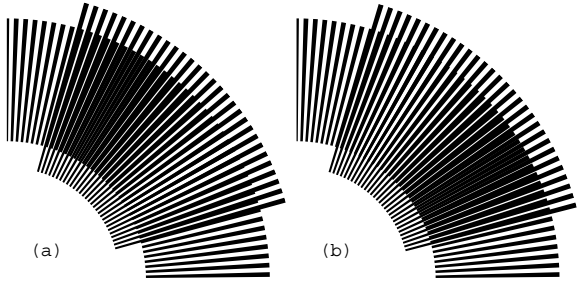

(a)  (b)

**Figure 14. Rotation of the revealing layer by 1 degree in the clockwise direction rotates the optical image by 30 degrees in the same direction [eps (a)], [png (a)], [eps (b)], [png (b)]**

In the GIF animation of the superposition image of Figure 13 the revealing layer slowly rotates in the clockwise direction.

## 3.2. Superposition of circular patterns with radial curves

In circular periodic patterns curved radial lines can be constructed using the same reference sequences of inclination degrees as used in section 2.3 for curves of Figure 6. The inclination angle at any point of the radial curve corresponds to the angle between the curve and the axis of the radius passing through the current point. Thus inclination angle 0 corresponds to straight radial lines as shown in Figure 13. With the present notion of inclination angles for $\alpha_b$, $\alpha_r$, and $\alpha_m$, equations (2.6) and (2.20) are applicable for circular patterns without modifications.

Curves can be constructed incrementally with a constant radial increment equal to $\Delta r$. Figure 15 shows a segment of a curve, marked by a thick line, which has an inclination angle equal to $\alpha$.

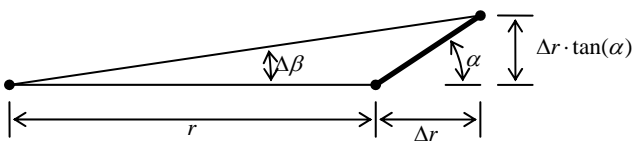

**Figure 15. Constructing a curve in a polar coordinate system with a desired inclination**

While constructing the curve, the current angular increment $\Delta\beta$ must be computed so as to respect the inclination angle $\alpha$:

$$\Delta\beta = \arctan\left(\frac{\Delta r \cdot \tan(\alpha)}{r + \Delta r}\right) \approx \frac{180°}{\pi \cdot r} \cdot \Delta r \cdot \tan(\alpha) \qquad (3.6)$$

Figure 16 shows a superposition of layers with curved radial lines. The inclination of curves of both layers follows an identical pattern corresponding to the following sequence of degrees (+35, –35, +35, –35, +35). Layer curves are iteratively constructed with increment pairs $(\Delta r, \Delta\beta)$ computed according to equation (3.6). Since the inclination patterns of both layers of Figure 16 are identical, the moiré curves also follow the same pattern.

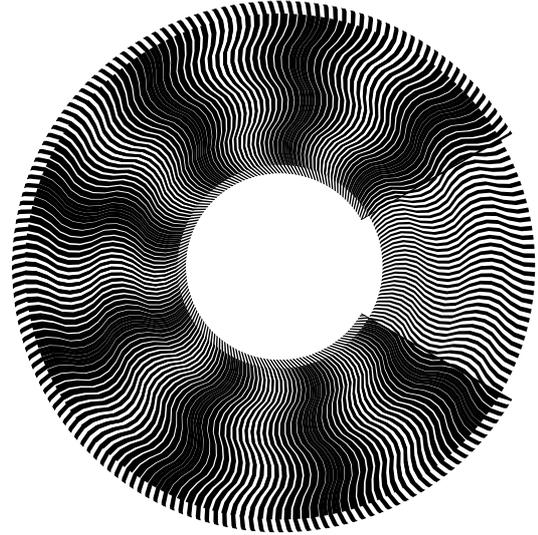

**Figure 16. Superposition of layers in a polar coordinate system with identical inclination patterns of curves corresponding to (+35, –35, +35, –35, +35); a portion of the revealing layer is cut away exposing the base layer in the background [eps], [png], multi-page [tif], [gif], small version [eps], [png]**

Similarly to examples of Figure 6, Figure 11, and Figure 12, where the same moiré pattern is obtained by superposing different pairs of layer patterns, the circular moiré pattern of Figure 16 can be analogously obtained by superposing other pairs of circular layer patterns. Taking into account equations (3.1), equations (2.6) and (2.20) can be rewritten as follows:

$$\tan \alpha_m = \frac{n_r \cdot \tan \alpha_r - n_b \cdot \tan \alpha_b}{n_r - n_b} \qquad (3.7)$$

$$\tan \alpha_r = \frac{n_b}{n_r} \cdot \tan \alpha_b + \left(1 - \frac{n_b}{n_r}\right) \cdot \tan \alpha_m \qquad (3.8)$$

Thanks to equation (3.8), other pairs of layer patterns can be created (see Figure 17) which produce the same superposition image as in Figure 16. In the first image (a) of Figure 17, the base layer lines are straight. In the second image (b), the base layer lines inclination pattern is reversed with respect to the moiré lines.

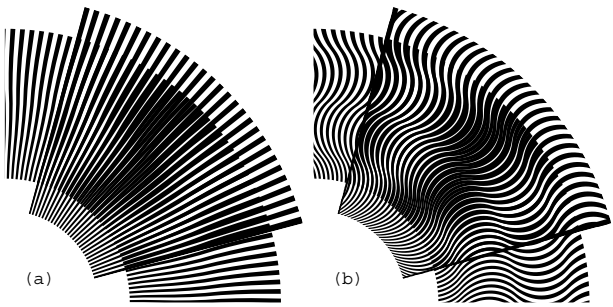

**Figure 17.** Superposition images with identical inclination pattern (+45, –45, +45, –45, +45) of moiré curves, where in one case the base layer comprise straight radial segments, and in the second case the base layer comprise curves which are the mirrored counterparts of the resulting moiré curves [eps (a)], [png (a)], [eps (b)], [png (b)]

We provide an animation, where the moiré curves of the superposition image are always the same, but the inclination pattern of the base layer curves gradually alternates between the following two mirror patterns (+45, –45, +45, –45, +45), and (–45, +45, –45, +45, –45) [eps], [tif], [gif]. For each instance of the animation, the revealing layer lines are computed according to equation (3.8) in order to constantly maintain the same moiré pattern.

Equations (3.4) and (3.3) remain valid for patterns with curved radial lines. In Figure 16 there are 180 curves in the revealing layer and 171 curves in the base layer. Therefore optical speedup factor according to equation (3.4) is equal to 20, and the number of moiré curves according to equation (3.3) is equal to 9, as seen in the superposition image of Figure 16.

## 4. Conclusions

We redeveloped the most important formulas for computing the periods, inclination angles of moiré patterns, and the velocities of optical shapes.

Instead of using the well known equations, we represent the case of inclined patterns such that equations (2.2), (2.3), and (2.4) for linear patterns and their counterparts (3.3), (3.5), and (3.4) for circular patterns, remain valid in their simple forms. In our equations, the *p* values represent the periods along the axis of the movement of the revealing layer.

In section 2.3.2 we compared our formulas with the formulas known in the literature.

## 6. Links

| | |
|---|---|
| 070212 | Random moiré [CH], [US] |
| 070227 | Random line moiré [CH], [US] |
| 070306 | Periodic line moiré patterns and optical speedup [CH], [US] |
| Web | [doc], [pdf], [htm], [htm (ms)] |
| Formats | [htm], with bitmaps [pdf], [doc], with vector graphics [pdf], [doc] |